\documentclass[acmtog,nonacm]{acmart}

\AtBeginDocument{%
  }

\usepackage{todonotes}
\usepackage{color}
\usepackage{rotating}
\usepackage{wrapfig}
\usepackage{subfig}
\graphicspath{{figures/}}
\usepackage{arydshln}
\usepackage{multirow} 
\usepackage{amsmath}
\usepackage{graphicx}
\usepackage[T1]{fontenc}

\definecolor{olive}{rgb}{0.5, 0.5, 0.0}

\usepackage{xspace}
\newcommand*{\eg}{e.g.\@\xspace}
\newcommand*{\ie}{i.e.\@\xspace}

\newcommand{\ssection}[1]{\noindent{\textbf{#1}}}

\newcommand{\customdance}{\textsc{CustomDance}}

\begin{document}

\title{CustomDance: Customized 3D Dance Generation with Coarse-to-Fine Human-Centered Interactive Control}

\author{Xulong Tang}
\authornote{Equal Contribution.}
\affiliation{%
  \institution{University of Texas at Dallas}
  \city{Richardson}
  \state{Texas}
  \country{USA}
}

\author{Kaixing Yang}
\authornotemark[1]
\authornote{Project Leader.}
\affiliation{%
  \institution{MalouTech Inc}
  \country{USA}
}

\author{Xiaohu Guo}
\affiliation{%
  \institution{University of Texas at Dallas}
  \city{Richardson}
  \state{Texas}
  \country{USA}
}

\author{Balakrishnan Prabhakaran}
\affiliation{%
  \institution{University at Albany}
  \city{Albany}
  \state{New York}
  \country{USA}
}

\author{Rawan Alghofaili}
\affiliation{%
  \institution{University of Texas at Dallas}
  \city{Richardson}
  \state{Texas}
  \country{USA}
}

\begin{abstract}
With the rise of AI-generated content (AIGC) and advanced techniques for 3D human representation, the task of generating 3D dance movements has become an exciting area of research. Despite significant advancements, current methods often fail to provide comprehensive and distinct control over various multimodal inputs from users, such as music or specific descriptions of desired movements. As a result, the generated motions may be statistically plausible and technically correct, but they often lack depth, expressiveness, and alignment with the user's creative vision. To address this issue, we present \customdance, a coarse-to-fine interactive system designed for customized 3D dance generation.
Inspired by the workflows of expert choreographers, \customdance~introduces a novel paradigm to AI-assisted choreography through three interconnected stages. First, a multimodal Large Language Model (MLLM) analyzes the music and a high-level text prompt to identify key temporal anchors and creative cues for the piece. Next, for each anchor, a multimodal retriever suggests high-quality motion clips from a dance library based on local music and text, empowering the user with concrete and predictable options. Finally, a custom music-conditioned diffusion in-painter seamlessly connects the selected phrases, allowing for iterative, user-guided refinement of the final composition, supported by visualizations of motion dynamics. Our evaluations demonstrate that \customdance~not only highlights the significant creative utility and empowering potential of our AI-assisted choreography paradigm, but also outperforms competitive baselines across quantitative and qualitative comparisons.
Project page: \url{https://github.com/XulongT/CustomDance}.

\end{abstract}


\begin{teaserfigure}
  \centering    
    \vspace{-2mm}
  \includegraphics[width=\textwidth]{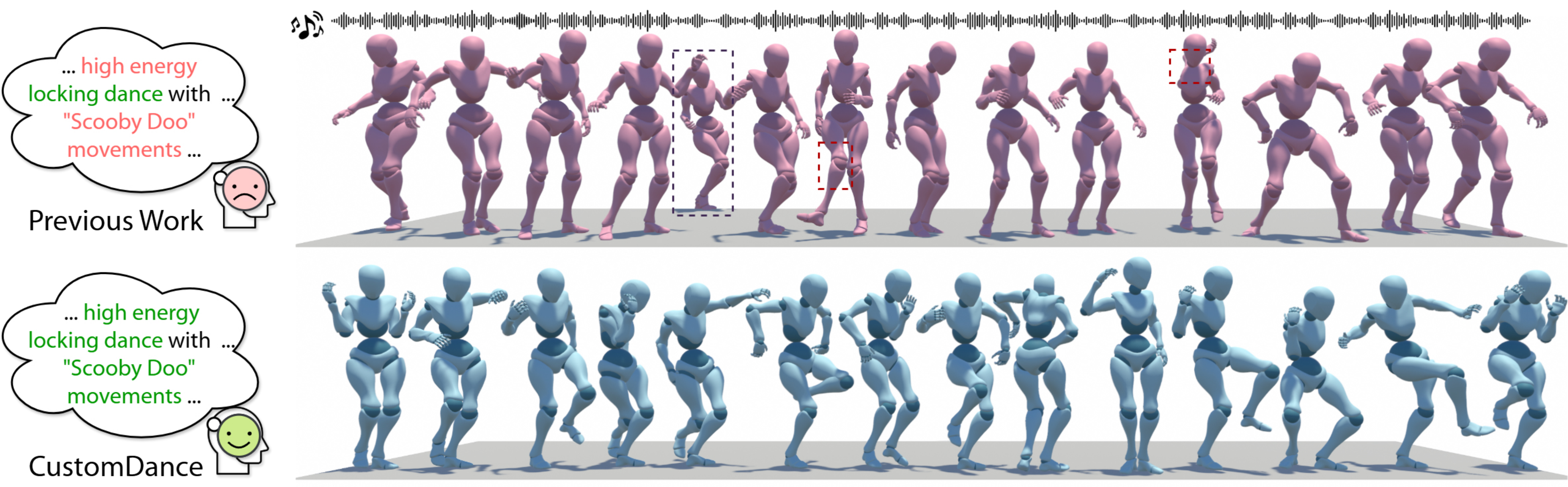}
  \vspace{-5mm}
  \caption{Users provide music and choreography preferences. Prior dance generation methods often fail to satisfy these preferences and frequently exhibit motion artifacts, such as root teleportation (purple box) and abnormal joint rotations (red box). \textbf{CustomDance} is a human-in-the-loop choreography system that generates 3D dances with substantially higher fidelity to user intent and greater artistic expressiveness.}
  \label{fig:teaser}
\end{teaserfigure}

\maketitle

\section{Introduction}
Dance is a powerful expression of human culture that transcends boundaries. Dancers convey both emotion and narrative intent by moving to the beat and melody, showcasing the power and beauty of human movement~\citep{butterworth2004teaching,blom1988moment}. Recent advancements in AI-generated content (AIGC)~\citep{yang2024beatdance,jia2026bitdiff,yang2026tokendance,zhang2026streamtalk,zhang2026personagesture} and 3D human representation techniques~\citep{loper2023smpl} have opened up exciting opportunities for automating the creation of 3D dance choreography. This is a timely and significant area for research, as such technology can benefit creators in fields like virtual entertainment, game development, and digital art. However, dance creation is inherently subjective and abstract; the same piece of music can yield multiple valid choreographic interpretations. This ambiguity presents a key challenge for 3D dance generation: how to produce choreography that accurately reflects a user's creative intent rather than resorting to a generic performance~\citep{huang2021choreography}.

Most existing approaches to 3D dance generation are purely music-driven, treating music as the primary conditioning signal for motion synthesis. These approaches employ a variety of generative models such as Generative Adversarial Networks (GANs)~\citep{yang2024cohedancers,sun2019deep,huang2021choreography}, Auto-Regressive (AR) models~\citep{yang2024codancers,yang2025matchdance,siyao2022bailando,siyao2023bailando++}, and Diffusion models~\citep{yang2026mace,yang2025flowerdance,tseng2023edge,li2024lodge}. However, music, as a largely fixed input, offers limited avenues for users to articulate their nuanced creative intent. To afford greater user control, subsequent works incorporate additional conditioning inputs, such as motion~\citep{tseng2023edge,liu2025gcdance}, genre~\citep{yang2025megadance,li2023finedance}, and text~\citep{gong2023tm2d,gupta2025mdd}. Nevertheless, these methods predominantly adopt an end-to-end paradigm that does not explicitly model user interaction or human-in-the-loop refinement. 
This design makes them poorly suited for iterative and expressive dance creation in real-world creative workflows (Fig.~\ref{fig:teaser}). 

Existing interactive 3D human–motion systems still lack comprehensive multimodal control of dance choreography.
Prior works have explored performance-driven control~\citep{ishigaki2009performance,liang2009performance}, example-based composition~\citep{calvert1991composition,arikan2002interactive}, sketch-based interfaces~\citep{thorne2004motion,choi2016sketchimo, davis2015drawing}, and extended reality (XR)-based manipulation~\citep{garcia2019spatial, zhou2024timetunnel}. These systems typically handle only short, simple motions and seldom consider musical structure or choreographic semantics. This makes them ill-suited for creating long, stylistically rich dances. Other work instead targets choreography directly, incorporating generative models or virtual reality (VR) environments into the choreographic workflow~\citep{liu2024dancegen, han2025choreocraft}. However, these tools still provide only limited control, and users cannot specify the alignment of key dance phrases with the music. Users are also unable to control global dance style, phrase-level semantics, and body-part–level motion patterns, which are critical for fine-grained personalization. As a result, the authored dances often appear generic and fall short of the user’s creative vision.

To bridge these gaps, we propose \customdance, a novel coarse-to-fine human-in-the-loop interactive system for customized 3D dance generation. We draw from prior dance and choreography work to implement a multi-stage, music-centered workflow~\citep{stevens2003choreographic}: First, during \emph{music analysis and phrasing}, choreographers listen to the music and segment it into musical phrases that define the structure of the choreography~\citep{blom1982intimate,smith2014dance}. Second, during \emph{movement exploration}, choreographers explore and select movements for each phrase based on the desired style and expressive emphasis~\citep{butterworth2004teaching}. Third, through \emph{structuring and refinement}, they assemble the full dance and iteratively refine transitions and local or fine-grained details through recombination and adjustment~\citep{gavish2020thinking}.

To mirror choreographers' real-world workflow, \customdance~ consists of three interconnected stages: In the \textbf{Choreographic Motif Planning} stage, \customdance~ leverages a multimodal large language model (MLLM)~\citep{yin2024survey} to analyze the input music along with the user’s intent (\eg, "a locking dance") to generate temporal anchors and corresponding creativity cues (\eg, "low frequency intro") that describe the music clip surrounding these anchors and suggest what to insert there (\eg, subtle chest isolations). The system then initializes phrase-level authoring slots around these anchors. Thereafter, the \textbf{Dance Phrase Generation} stage allows the user to assign each empty phrase slot a dance phrase. This is done by retrieving a top-10 list of candidate dance phrases conditioned on the slot's music clip, additional user-provided text prompts (\eg, "long reaches, hand claps"), and body-part controls (\eg, selecting the right-arm joint group and increasing its intensity weight). Finally, during the \textbf{Completion and Refinement} stage, users can fill any empty gaps between phrases via automated inpainting and repair or refine undesirable movements. \customdance~ not only effectively leverages a broader range of multimodal information but also strategically assigns each modality to a specialized stage with a distinct role. This comprehensive design ensures high fidelity to the user's creative intent while elevating the final quality and expressiveness of the generated 3D dance.


In summary, our main contributions are:
(1) We present \customdance, a coarse-to-fine, human-in-the-loop interactive system for customized 3D dance generation, inspired by the creative workflow of real-world expert choreographers.
(2) We design a modular choreography authoring workflow composed of three interactive stages: MLLM-based anchor and cue generation, user-controllable multimodal dance phrase retrieval, and music-conditioned diffusion inpainting with interactive refinement.
(3) We conduct extensive experiments that demonstrate the practical effectiveness and creative flexibility of our AI-assisted choreography system, and show consistent improvements over competitive baselines in both quantitative metrics and qualitative comparisons.

\section{Related Work}

\subsection{Dance Motion Generation}
Music and dance are deeply intertwined, and recent progress in music-to-dance generation has largely centered on 3D motion~\citep{fan2025align}. Broadly, existing methods fall into three families: GAN-based, autoregressive, and diffusion-based models.
In GAN-based models, generators synthesize motion from music while discriminators provide adversarial feedback~\citep{yang2024cohedancers,sun2019deep,huang2021choreography}. Autoregressive models typically adopt a two-stage pipeline, first curating choreographic units and then modeling music-conditioned distributions over these units~\citep{siyao2023bailando++,siyao2024duolando,yang2025megadance}.
Diffusion-based models corrupt motion with noise and train denoising networks to iteratively recover sequences conditioned on music, enabling diverse and temporally coherent dances~\citep{tseng2023edge,li2023finedance,li2024lodge++,li2025music,yang2025flowerdance,nguyen2025egomusic}. In addition, some retrieval-based methods~\citep{chen2021choreomaster, huang2025motionrag} compose and refine dances from motion libraries to improve motion fidelity. However, using music as the only input for generation offers limited avenues for users to articulate their nuanced creative intent. To afford greater user control, some works incorporate additional conditioning inputs, such as human motion~\citep{tseng2023edge,liu2025gcdance}, dance genre~\citep{yang2025megadance,li2023finedance}, physical or cognitive intensity~\citep{tang2026personalized}, keyframe constraints~\citep{yang2023keyframe}, or text and multimodal controls~\citep{gong2023tm2d,gupta2025mdd,gupta2025unified}. Thus, existing methods typically treat their inputs as static conditions in end-to-end pipelines, with limited support for interactive authoring. Our work instead supports a human-in-the-loop choreographic workflow, allowing for user-specified creative intent to guide generation.

\begin{figure*}[t]
    \centering
    \includegraphics[width=0.95\linewidth]{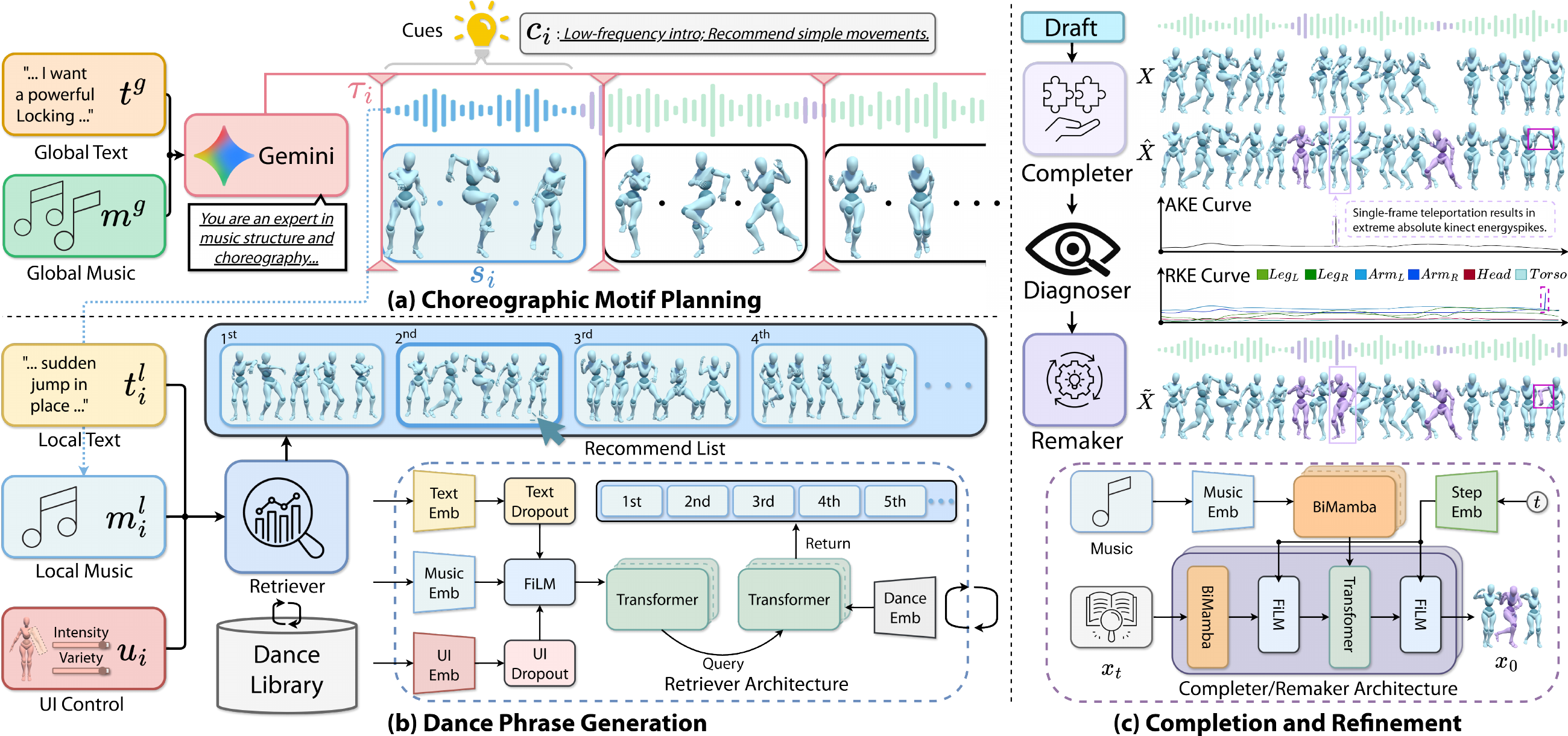}
    \caption{
    Overview of \customdance~.
    }
    \label{fig:overview}
\end{figure*}

\subsection{Interactive 3D Motion Authoring} 

To enable creators to design 3D movements without directly manipulating low-level joint trajectories, interactive 3D motion authoring systems and control operators have been widely studied in computer graphics and human-computer interaction~\citep{gou2025control}. Broadly, prior systems can be grouped into four categories: performance-driven control, example-based composition, sketch-based interfaces, and immersive XR (VR/AR) interfaces. Performance-driven systems map users' physical movements to virtual characters in real time~\citep{ishigaki2009performance,liang2009performance}. Example-based composition systems build motions from a library by searching, stitching, and retiming clips according to user-specified spatial or temporal constraints~\citep{calvert1991composition,arikan2002interactive}. Sketch-based interfaces interpret 2D or 3D drawings as motion cues and map sketches to parameterized motion primitives~\citep{thorne2004motion,choi2016sketchimo,davis2015drawing}. Finally, immersive XR interfaces enable in-situ motion authoring through embodied interaction, supporting spatial sketching, spatio-temporal manipulation, and context-aware authoring~\citep{garcia2019spatial,zhou2024timetunnel,alghofaili2023,kim2024,anderson2013,ye2020}.
These systems effectively facilitate motion control and basic character animation. However, they primarily accommodate short and relatively simple movements, and often neglect musical structures or choreography semantics, making them challenging to apply to longer and more stylistically intricate dance sequences.
Recent motion authoring systems have moved toward dance choreography by incorporating generative models and immersive environments into creative workflows~\citep{liu2024dancegen,han2025choreocraft}. However, these systems do not yet provide an interaction paradigm that matches real-world choreographic practice, where creators iteratively plan, compose, and refine dances in response to music.



\section{Overview}
\begin{figure*}[t]
  \centering
  \begin{tabular}{@{}cc@{}}
    \includegraphics[width=0.48\textwidth]{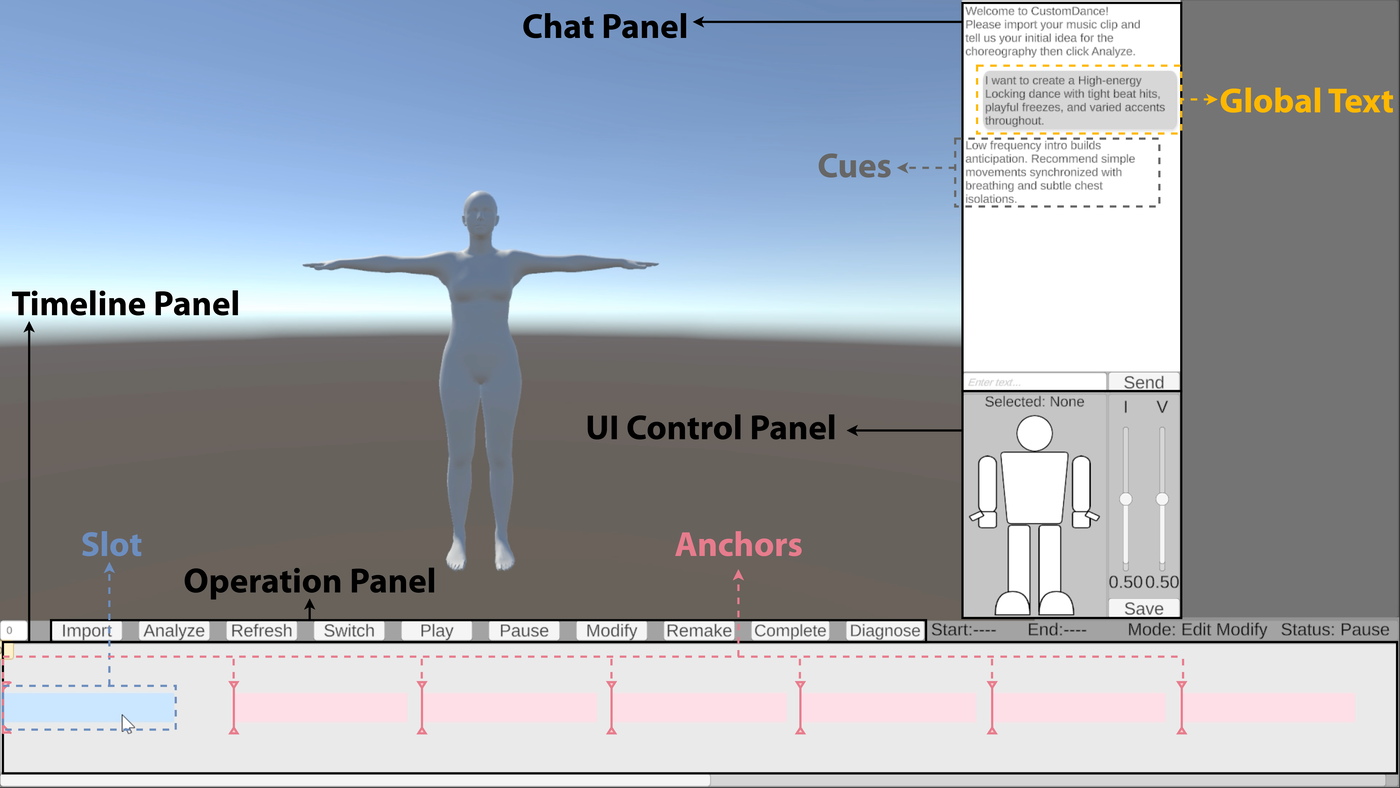} &
    \includegraphics[width=0.48\textwidth]{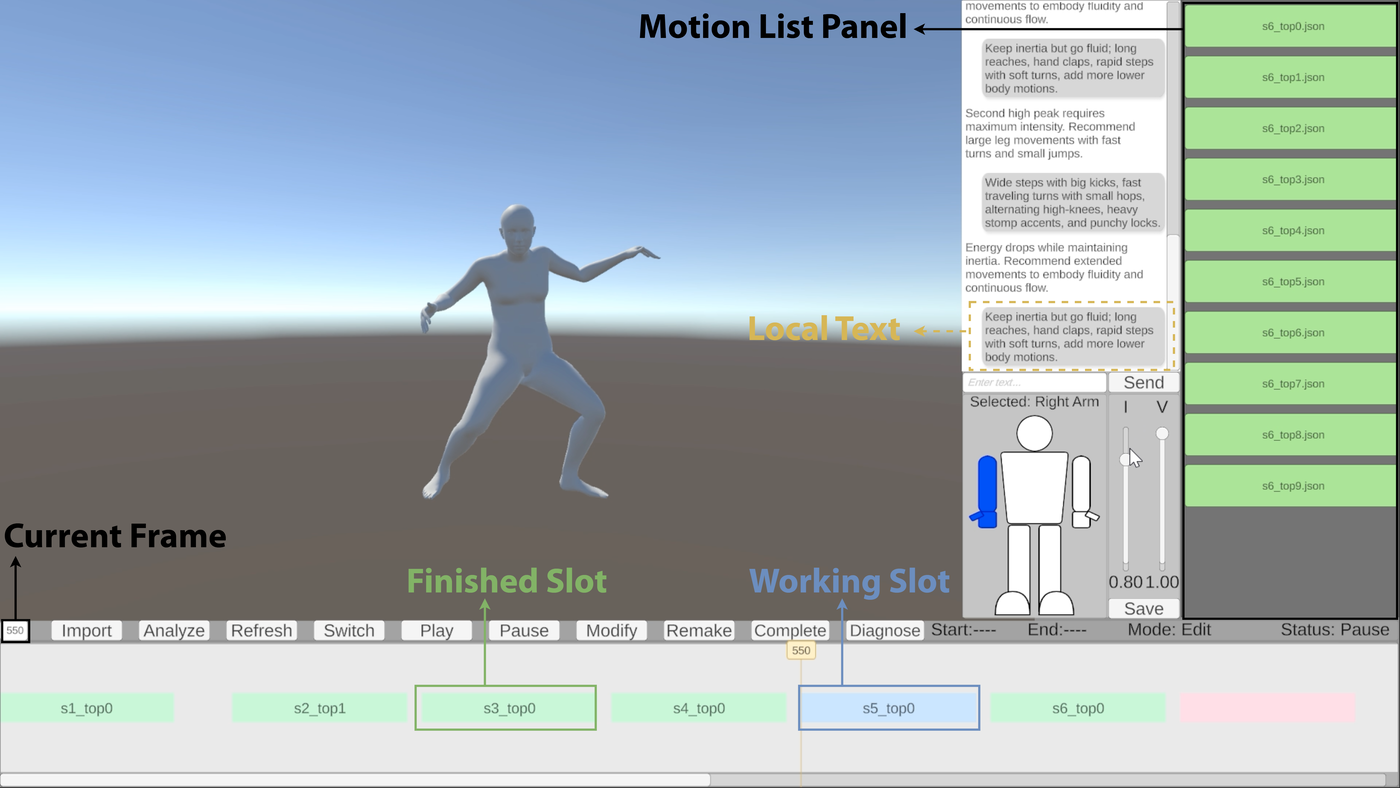} \\
    \small (a) Choreographic Motif Planning &
    \small (b) Dance Phrase Generation \\
    \includegraphics[width=0.48\textwidth]{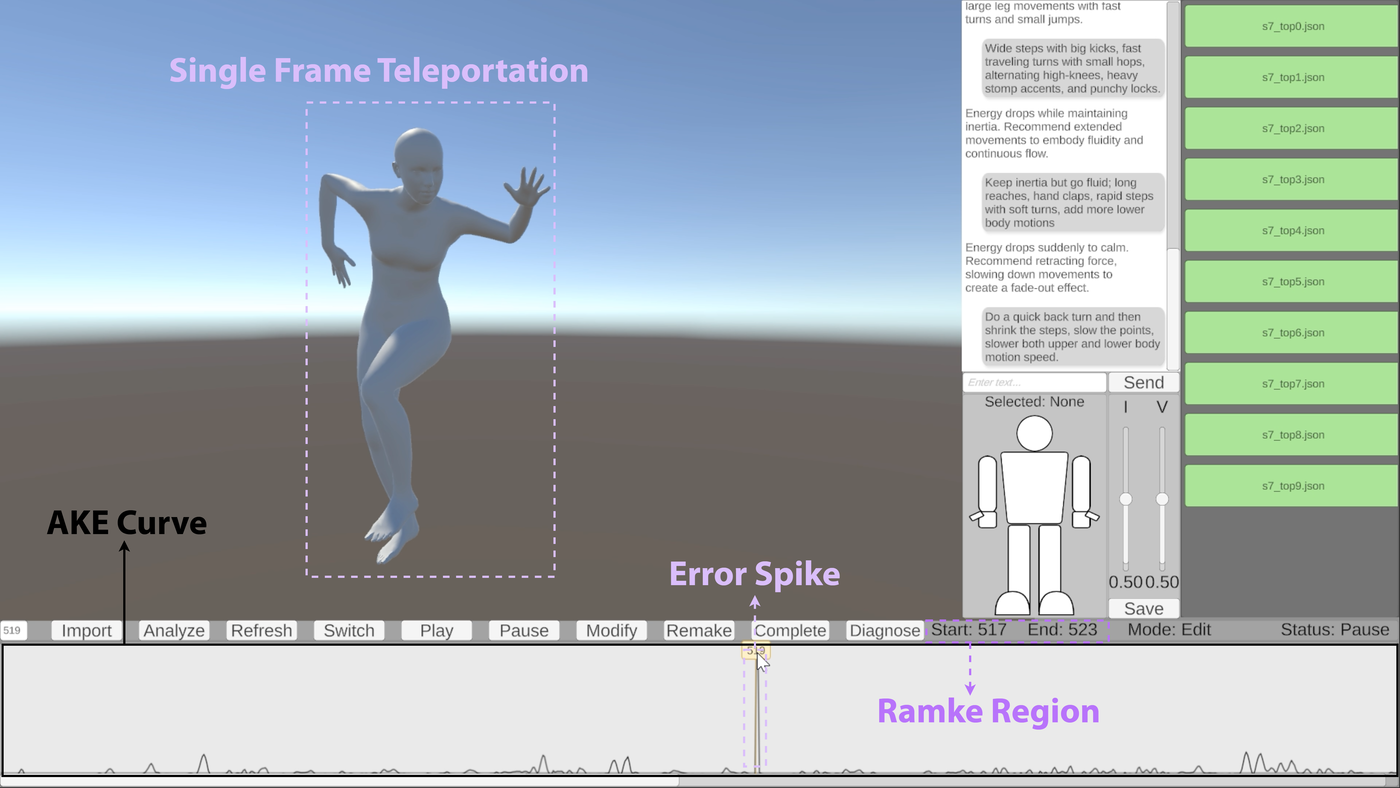} &
    \includegraphics[width=0.48\textwidth]{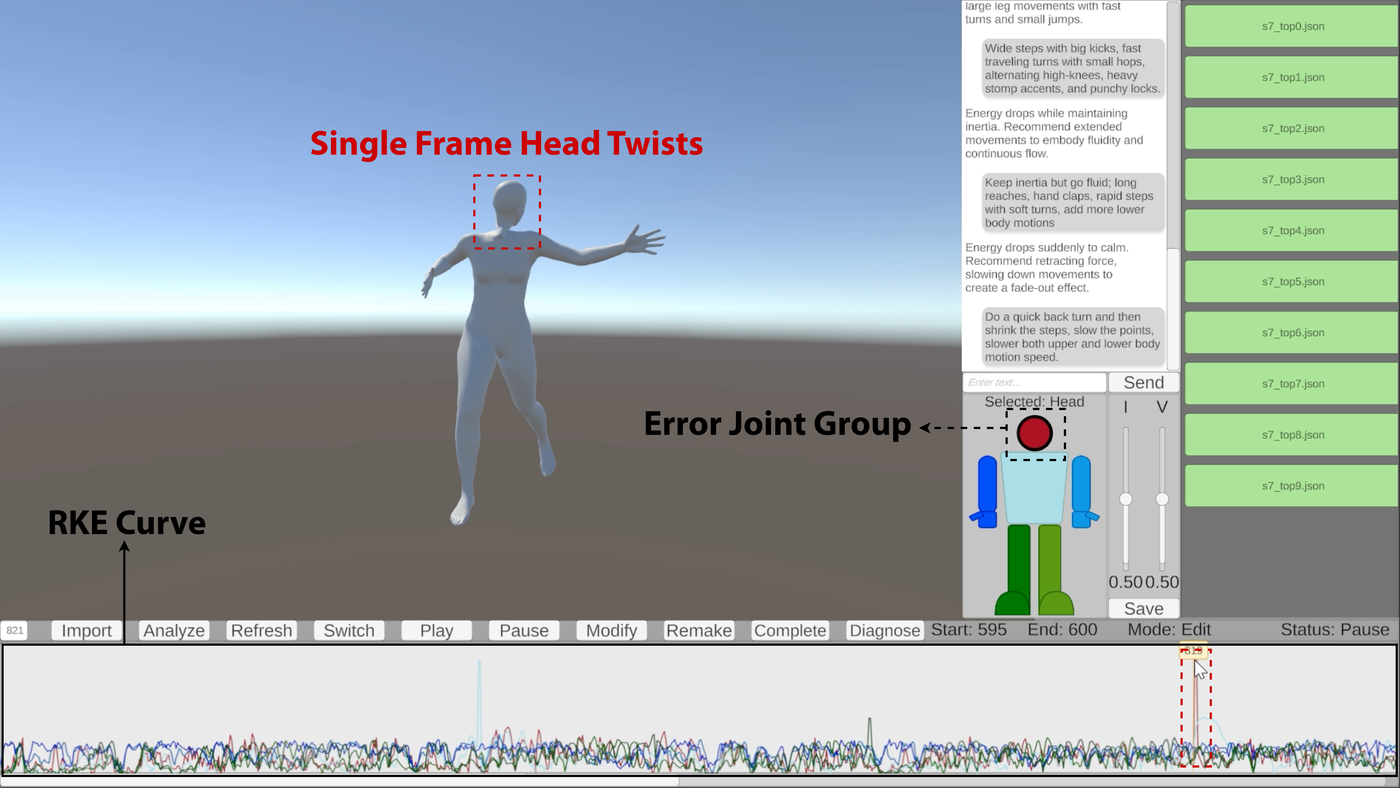} \\
    \small (c) Completion and Refinement (AKE) &
    \small (d) Completion and Refinement (RKE)
  \end{tabular}
  \caption[System overview]{System overview of the interactive choreography pipeline.}
  \label{fig:ui}
\end{figure*}

\customdance~ is a human-in-the-loop, coarse-to-fine choreography system inspired by professional dance-making workflows~\citep{stevens2003choreographic} (Fig.~\ref{fig:overview}). Through user inputs across different authoring stages, it produces a dance sequence $\tilde{X}$ in SMPL representation~\citep{loper2023smpl}. \customdance~ has a three-stage pipeline: 

\ssection{Choreographic Motif Planning.} 
The user imports a global music track $m^{g}$ (\eg, a locking track) and provides a global textual description $t^{g}$ (``\dots I want a powerful Locking dance\dots'') in the chat panel labeled as ``Global Text'' in Fig.~\ref{fig:ui}a (Fig.~\ref{fig:overview}a).
An MLLM (Gemini~\citep{team2023gemini}) analyzes $(m^{g}, t^{g})$ and outputs a set of temporal anchors $\{\tau_i\}$ with corresponding creativity cues $\{c_i\}$ (``Low-frequency intro; Recommend simple movements''). 
\customdance~ converts anchors into phrase-level authoring slots and initializes them on the timeline as shown in Fig.~\ref{fig:ui}a; when the user selects a slot $s_i$, the system displays its suggested cue to help the user edit it into a more desirable local description $t^{l}_{i}$. If the user provides no local text description, it defaults to $c_i$.

\ssection{Dance Phrase Generation.} 
For each selected working slot $s_i$, the user specifies a local text prompt $t^{l}_{i}$ (``\dots sudden jump in place\dots'') and sets the joint-group intensity/variety controls $u_i$ (or keeps the default settings at 0.5) in the UI control panel shown in Fig.~\ref{fig:ui}a (Fig.~\ref{fig:overview}b).
Conditioned on the slot music clip $m^{l}_{i}$, a multimodal retriever then queries a curated dance library using $(m^{l}_{i}, t^{l}_{i}, u_i)$ and returns the Top-$K$ candidate phrases to the ``Motion List'' panel for user selection, enabling iterative slot filling with previewable options. For example, the user can enter ``add long reaches'' into the chat panel and increase the right-arm intensity value (Fig.~\ref{fig:ui}b). The user can select any slot and repeat these steps to create the choreography draft $X$, as shown at the bottom of Fig.~\ref{fig:ui}b.


\ssection{Completion and Refinement.} 
After filling all desired slots to produce the choreography draft $X$, the user can click the ``Complete'' button to trigger a music-conditioned diffusion model that inpaints the gaps between selected slots, resulting in a gapless choreography $\hat{X}$~\citep{liu2025gcdance,tseng2023edge} (Fig.~\ref{fig:overview}c). 
By clicking the ``Diagnostic'' button, \customdance~ diagnoses the choreography draft by visualizing absolute and relative kinetic-energy curves over time~\citep{li2021ai}, which help users quickly locate abnormal intervals such as single-frame teleportation or abnormal limb twists.
The user can then select these regions on the timeline (highlighted in rectangles on $\hat{X}$ in Fig.~\ref{fig:overview}c) and invoke the Remaker for localized re-synthesis. The remake can target either the full body or a joint group until the choreography matches the intended style and quality, producing $\tilde{X}$, as shown in Fig.~\ref{fig:ui}c--d.

\section{Choreographic Motif Planning}
The user provides a global music track $m^g$ and a global choreography description text $t^g$ to define the choreographic intent. We feed $(m^g, t^g)$ into an MLLM (Gemini) to analyze musical structure and output a set of temporal anchors $A=\{\tau_i\}_{i=1}^{N}$ with corresponding creative cues $C=\{c_i\}_{i=1}^{N}$, where $N$ is the number of suggested anchors. Each anchor $\tau_i$ denotes a suggested onset for phrase-level authoring. \customdance~maps each anchor to a retrieval slot $S=\{s_i\}_{i=1}^{N}$ as $s_i=[\tau_i,\tau_i+\Delta_{\text{slot}}]$, where $\Delta_{\text{slot}}$ is the slot-duration parameter shared by the retrieval library, preview interface, and timeline editor. In our implementation, we set $\Delta_{\text{slot}}=4$\,s. Each cue $c_i$ describes the local music around slot $s_i$ and recommends a suitable movement.

We guide Gemini with a structured prompt that asks it to analyze coarse musical sections, tempo, energy changes, and rhythmically salient events while interpreting $t^g$ as the long-term choreographic goal. It then returns JSON-formatted anchor--cue pairs, where anchors respect a minimum temporal separation and cues are written as short imperative phrases for downstream reuse. We sort, deduplicate, clamp, and convert the returned anchors into frame indices before constructing slots. The system visualizes $(\tau_i, s_i, c_i)$ on the timeline, allowing users to review anchors and cues. Generated slots may overlap. Selecting one temporarily invalidates conflicting slots, which are restored when the user undoes or changes the selection.

\section{Dance Phrase Generation}
After Choreographic Motif Planning, the user selects a slot $s_i$ and populates it via multimodal retrieval from a curated dance library. We use retrieval to provide realistic and stylistically coherent human-motion phrases, enabling direct and interpretable user choices. Generative modeling is reserved for gap filling and localized repair.

The query is triple-modal: $q_i=(m_i^l, t_i^l, u_i)$, consisting of local music $m_i^l$, local text $t_i^l$, and joint-group UI controls $u_i$.

\ssection{Local music.} $m_i^l$ is the $\Delta_{\text{slot}}$-second audio segment aligned with $s_i$. We encode $m^l_i$ using Librosa features~\citep{mcfee2015librosa} (\ie, MFCCs with derivatives, CQT chroma, tempogram, and onset/beat features) to capture timbre, harmony, and rhythm characteristics.

\ssection{Local text.} $t_i^l$ specifies phrase-level intent (\eg, style, semantics). If the user does not provide $t_i^l$, the system uses the corresponding creative cue $c_i$ as a default description for that slot. We encode $t_i^l$ into a 512-D vector with a CLIP text encoder~\citep{radford2021learning}.

\ssection{UI Controls.}
$u_i \in \mathbb{R}^{6 \times 2}$ controls six joint groups (head, torso, left/right arms, and left/right legs). For each group, \emph{intensity} specifies motion magnitude and spatial extent, while \emph{variety} specifies movement diversity relative to repetitive motion patterns. Prior work has shown that this \emph{intensity} setting correlates with physical exertion, while \emph{variety} correlates with choreography memorability and learning difficulty~\citep{tang2026personalized}. The normalized control values are flattened and encoded into a 512-D vector using an MLP.

\subsection{Retrieval Method}
We adopt a contrastively trained trimodal Retriever (Fig.~\ref{fig:overview}). On the query side, concatenated local-text and UI-control embeddings form a user-condition vector that modulates the local-music embedding through FiLM~\citep{perez2018film}. A Transformer then produces the query embedding. On the library side, each $\Delta_{\text{slot}}$-second SMPL dance phrase is encoded by a motion encoder and a Transformer into a phrase embedding; these library embeddings are precomputed offline. At runtime, we rank phrases by cosine similarity to the query embedding and return the Top-$K$ candidates ($K=10$).

\ssection{Training.} To improve robustness to incomplete user inputs, we apply modality dropout during training: with probability $\lambda_{\mathrm{drop}}=0.2$, we replace either the text or UI modality with a modality-specific learnable token. This dropout models missing modalities, while the runtime default cue/text and slider values only initialize usable queries. We train the retriever with an InfoNCE loss that aligns each query embedding with its paired dance phrase. Full objective and implementation details are provided in the supplementary material.

\section{Completion and Refinement}
After generating dance phrases, the user receives a draft sequence $X$ with selected slots filled in. \customdance~then refines this draft through gap filling, artifact diagnosis, and localized repair. The \emph{Completer} inpaints unfilled temporal gaps to produce a seamless draft $\hat{X}$, while the \emph{Diagnoser} and \emph{Remaker} help users locate and resynthesize occasional artifacts (\eg, single-frame root jumps or twisted limbs), producing the final choreography $\tilde{X}$.

\subsection{Gap Filling and Dance Completion}
The draft $X$ contains selected phrase intervals and unassigned gaps.
We formulate gap filling as inpainting under our diffusion backbone.
Specifically, we construct a spatio-temporal mask $\mathbf{m}$ that marks retrieved phrase frames as \emph{known} and gap frames as \emph{unknown}.
During reverse diffusion, we keep all known phrase frames fixed and inpaint only the unknown (gap) frames at each denoising step.
This yields a continuous full-length choreography $\hat{X}$ that smoothly connects the retrieved phrases under aligned music conditioning.

\subsection{Artifact Detection and Repair}

To help the user efficiently locate subtle artifacts, we compute lightweight kinetic-energy proxies from per-joint linear and angular motion between consecutive poses~\citep{Onuma2008FMDistance, Yamane2009}. Let $p$ be the current pose and $p'$ its preceding pose in the synthesized sequence, and let $\Delta_{Frame}$ denote the transition time (seconds). 
For joint $j$, we denote its 3D position by $\mathbf{o}_{p,j}\in\mathbb{R}^3$ and its rotation by $\mathbf{R}_{p,j}\in SO(3)$.

\ssection{Linear Velocity.} The linear velocity proxy is
\setlength{\abovedisplayskip}{0.5pt}
\setlength{\abovedisplayshortskip}{0.5pt}
\setlength{\belowdisplayskip}{0.5pt}
\setlength{\belowdisplayshortskip}{0.5pt}
\begin{equation}
\mathbf{v}_{p,j}=\frac{\mathbf{o}_{p,j}-\mathbf{o}_{p',j}}{\Delta_{Frame}} 
\end{equation}
\ssection{Angular Velocity.} The angular velocity proxy is computed on $SO(3)$ via the logarithm map~\citep{grassia1998practical}:
\begin{equation}
\label{eq:omega}
\boldsymbol{\omega}_{p,j}=\frac{\mathrm{log}\!\left(\mathbf{R}_{p',j}^{-1}\mathbf{R}_{p,j}\right)}{\Delta_{Frame}} 
\end{equation}

\ssection{Diagnostics.} Using these kinematics-based proxies (omitting mass and inertia constants, which only introduce a global scaling), we define absolute and relative energy curves:
\begin{equation}
E_{\text{abs}}(p)=\frac{1}{|G_{\text{torso}}|}\sum_{j\in G_{\text{torso}}}\|\mathbf{v}_{p,j}\|^2
\end{equation}
\begin{equation}
E_{\text{rel}}(p,g)=\frac{1}{|G_g|}\sum_{j\in G_g}\|\boldsymbol{\omega}_{p,j}\|^2 
\end{equation}
$E_{\text{abs}}$ (AKE) captures global translation spikes (\eg, root jumps), while $E_{\text{rel}}$ (RKE) captures abrupt local rotational changes within a joint group (\eg, limb twists). 
We use six joint groups $\{G_g\}_{g=1}^{6}$: \emph{head}, \emph{torso}, \emph{left/right arms}, and \emph{left/right legs}. 
The system overlays curve peaks on the timeline to help users inspect possible artifacts, since high-energy accents may also appear in intentional dance movements. Users inspect the flagged intervals and invoke the Remaker when localized re-synthesis is desired, as shown in Fig.~\ref{fig:ui}c--d.

\ssection{Repair.} Repair is implemented via masked denoising (inpainting) under the same diffusion backbone: we construct a spatio-temporal mask that marks the selected joints within the chosen window as \emph{unknown} and keeps all other joints/frames as \emph{known}.
This mask can cover the full body (for global remake) or a single joint group (for partial remake), enabling resynthesis of only the targeted motion while preserving the remaining joint groups and the overall phrasing.
Compared with full-gap completion, the Remaker solves a more constrained local inpainting problem: it operates within a user-selected window while keeping surrounding frames and unselected joints fixed.
The output is the refined choreography $\tilde{X}$.

\subsection{Music-to-Dance Generation Model}
We use a DDIM-based music-to-dance model to synthesize and edit motion in \customdance~\citep{ho2020denoising,song2020denoising}. A motion sequence $x$ is represented in SMPL parameters~\citep{loper2023smpl}. During training, we corrupt clean motion with Gaussian noise and train a network $\hat{x}_\theta(x_t,t,f_m)$ to predict the clean sequence $x_0$ from noisy motion $x_t$, diffusion timestep $t$, and paired music features $f_m$. We use an $x_0$-prediction reconstruction loss and auxiliary kinematic losses on joint positions, joint velocities, and foot velocities following EDGE~\citep{tseng2023edge}. Full training objectives are provided in the supplementary material.

\noindent\textbf{Inference.}
At inference, we use DDIM sampling to progressively refine noise into a final dance sequence. We use classifier-free guidance (CFG): during training, we drop $f_m$ with probability $0.25$, and during inference we apply a guidance scale of $2.5$.

\noindent\textbf{Architecture.}
As shown in Fig.~\ref{fig:overview}c, our generator follows a BiMamba--Transformer hybrid architecture inspired by MEGADance~\citep{yang2025megadance}. BiMamba models local temporal dependencies and bidirectional information flow within music and motion streams, while Transformer layers model global cross-modal interactions between the noisy motion state and encoded music features. The generator predicts denoised SMPL motion $\hat{x}_\theta(x_t,t,f_m)$. We train the model only on fixed 8-second clips, while Mamba's inherent scalability supports longer sequence completion at inference.

\ssection{Editing.}
To support completion and localized refinement, we use masked denoising as in diffusion-based inpainting~\citep{tevet2022human,kim2023flame}. Given a constraint $x^{\text{known}}$ and a binary mask $\mathbf{m}$, each denoising step preserves known regions and resynthesizes unknown regions:
    \begin{equation}
\hat{x}_{t-1} := \mathbf{m} \circ q(x^{\text{known}}, t-1) + (1 - \mathbf{m}) \circ \hat{x}_{t-1}
\end{equation}
where $\circ$ denotes the Hadamard product. This masked operator enables both temporal gap filling and joint-level local repair without additional training.

\noindent
\section{Experiments}
\subsection{Experimental Setup}

\ssection{Dance Library.}
Our dance library combines two sources of 3D motion: (1) motion-capture data and (2) motion reconstructed through human mesh recovery (HMR), totaling approximately 10.7 hours across 16 fine-grained dance genres and five coarse dance-style classes. The motion-capture portion comprises 7.7 hours of professionally captured motion data at 30 fps from FineDance~\citep{li2023finedance}, the largest publicly available 3D motion-capture dataset for music-to-dance generation. For the HMR portion, inspired by prior work on internet dance-data curation~\citep{yang2026omnidance}, we carefully select approximately 3 hours of high-quality dance videos in which the dancer's full body remains visible and unobstructed throughout, with continuous shots and limited camera motion. These selection criteria facilitate high-quality 3D SMPL motion reconstruction using PromptHMR~\citep{wang2025prompthmr}. We then align the reconstructed motions with FineDance in coordinate origin, ground plane, and coordinate system, and map their annotations to the FineDance schema. Finally, we augment the combined 3D motion corpus using StableMotion~\citep{mu2025stablemotion}.

\ssection{Training Clips.}
For phrase retrieval, we segment each choreography into 4-second retrieval units at the configured slot duration, resulting in approximately 6,900 motion phrases. For each phrase, we compute normalized \emph{intensity} and \emph{variety} measures for joint groups using kinetic features from prior work~\citep{Onuma2008FMDistance}. For the diffusion-based generation model, we segment the augmented corpus into 8-second training clips, yielding approximately 3,400 motion sequences. Additional implementation details are provided in the supplementary material.



\ssection{Music Clips and Choreography Description.}
We select five 32-second music clips whose paired reference choreographies span the five coarse dance-style classes in FineDance. To avoid trivial retrieval, all phrases derived from these choreographies are excluded from the dance library. For each clip, we create a choreography description using a consistent template derived from the corresponding dance style. For Study 1, we also precompute the anchor and cue texts and keep them fixed across participants to reduce variability across study conditions. The templates, descriptions, and music clips are included in the supplementary material.

\ssection{Apparatus and Run-time Performance.}
\customdance~is implemented in Unity.
All user-study runtime measurements are collected on an Alienware Area-51 equipped with an Intel Core Ultra 9 285K CPU, 64GB RAM, and an NVIDIA RTX 5090 GPU.
In the user studies, timeline initialization for Choreographic Motif Planning takes 0.5~s on average per music excerpt; each Dance Phrase Generation retrieval takes 0.7~s on average; and in Completion and Refinement, each Remaker operation takes 1.9~s per 100 frames on average.


\subsection{Study 1: Ablation of the Authoring Experience}
\label{sec:study_ablation}

\subsubsection{Ablation Study Design}\mbox{}\par

\ssection{Participants.}
We recruited 25 participants (12 males and 13 females) with diverse experience in dance performance (0--10 years; $M=3.28$, $SD=2.82$) from the university's student and staff population.
The study received Institutional Review Board (IRB) approval, and all participants provided informed consent.

\ssection{Study Design.}
This study evaluates how the three stages of \customdance~contribute to the authoring workflow.
The full system includes motif planning, multimodal phrase retrieval, and Diagnoser-guided completion/refinement.
We compare it with ablated variants that remove one stage-specific capability while preserving a comparable timeline-based workflow.
These ablations test whether Stage 1 provides useful music-aware structure, whether Stage 2 helps users find suitable phrases efficiently, and whether Stage 3 helps users identify and repair local motion artifacts.
We report targeted component-level evaluations after this ablation study, with full protocols and additional results provided in the supplementary material.

We adopt a within-subject ablation study design where we evaluate five authoring interface conditions: (1) \textit{\customdance},
(2) \textit{w/o Anchors+Cues},
(3) \textit{w/o Retriever},
(4) \textit{w/o Diagnoser},
and (5) \textit{Baseline}.
In the \textit{w/o Anchors+Cues} condition, we replace MLLM-generated anchors with uniform slots of the same duration and remove anchor-specific cues.
We ablate anchors and cues together because each cue is generated from the local music context around its anchor; uniform slots preserve a comparable timeline workflow while removing the music-aware anchor--cue suggestions.
We also apply a 0.5-second overlap with a linear blend at the boundaries of the slots to ensure that the final output is smooth and maintains a consistent length of 32 seconds.
In the \textit{w/o Retriever} condition, the Top-$K$ ranking mechanism is disabled and the motion list is populated by randomly sampled library clips.
This condition keeps the library-based authoring workflow unchanged, but replaces ranked recommendations with unassisted browsing to evaluate the role of multimodal ranking.
In the \textit{w/o Diagnoser} condition, we remove the Diagnoser's AKE/RKE kinetic-curve visualization while keeping the Completer/Remaker enabled, testing whether the Diagnoser helps users locate intervals that require repair.
The \textit{Baseline} is an editor that lacks the modules from the other conditions as well as the Completer/Remaker.

\ssection{Procedure.}
Each participant completed five authoring tasks, one per interface condition.
For each task, participants were given a music clip and its dance choreography description and were asked to author a complete 32-second choreography under a fixed time budget (30 minutes per task, early submission allowed).
The order of authoring conditions was counterbalanced using a balanced Latin square to mitigate order effects.
Additionally, to ensure that the evaluation of each interface condition is independent of the assigned music clip, we cycled through all possible combinations of music clips and authoring interface pairings throughout the study.

Prior to completing the conditions, participants underwent a brief tutorial and a practice trial using a music clip that was not part of the evaluation.
Short breaks were provided between tasks to minimize fatigue.
After each condition, participants completed 7-point Likert ratings on:
\emph{Perceived Controllability} (“I could effectively control the motion style and intensity to match my intention.”),
\emph{Pose Satisfaction} (“I am satisfied with the quality and naturalness of the poses in the final dance.”),
and \emph{Description Fulfillment} (“The final dance matches the given description of the choreography.”).

We also tracked three objective measures during the study conditions: (i) the authoring time, (ii) the number of clip replacements made, and (iii) the count of artifacts in each exported 32-second dance. The artifact count is an offline objective counting metric based on thresholded kinematic anomalies, separate from the AKE/RKE curves used for interactive Diagnoser visualization; details are provided in the supplementary material.

\subsubsection{Authoring Experience Results}\mbox{}\par
\begin{figure*}[t]
  \centering
  \includegraphics[width=\linewidth]{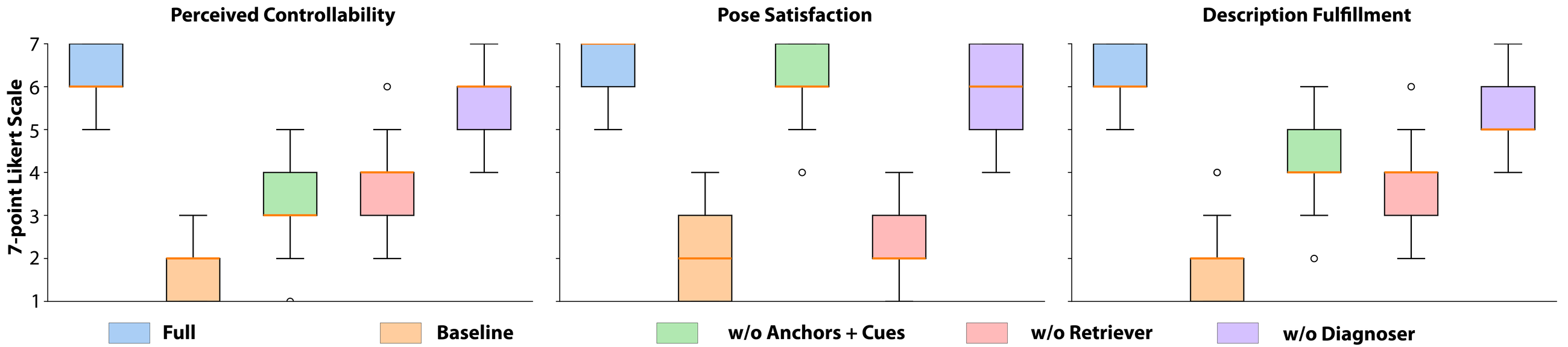}
  \caption[Task-level Likert ratings]{
  Task-level 7-point Likert ratings on Perceived Controllability, Pose Satisfaction,
  and Description Fulfillment across the five interface conditions.}
  \label{fig:taskrating}
\end{figure*}

Overall, the ablations show complementary benefits across the three stages.
Motif planning improves structure and description fulfillment.
Multimodal ranking reduces library-browsing effort and improves perceived controllability.
Diagnoser-guided refinement reduces unresolved artifacts and improves pose satisfaction.
Across all three subjective questions, \customdance~achieves the best ratings (Fig.~\ref{fig:taskrating}).
A Friedman test revealed a significant effect of interface condition on \emph{Perceived Controllability} ($\chi^2(4)=81.06$, $p<.001$, Kendall's $W=0.81$), \emph{Pose Satisfaction} ($\chi^2(4)=84.54$, $p<.001$, $W=0.85$), and \emph{Description Fulfillment} ($\chi^2(4)=88.17$, $p<.001$, $W=0.88$).

Wilcoxon signed-rank post-hoc tests with Holm correction show that \customdance~
significantly improves \emph{Pose Satisfaction} and \emph{Description Fulfillment} over the other conditions
($p_{\mathrm{Holm}}< .01$).
This suggests that motif planning provides phrase-level structure that better supports satisfying the choreography description.
Meanwhile, the Diagnoser enables targeted repair of kinematic issues, reducing unresolved artifacts and improving perceived pose quality.
For \emph{Perceived Controllability}, \customdance~ significantly outperforms the \emph{w/o Retriever} condition
($p_{\mathrm{Holm}}< .05$).
The Retriever reduces manual browsing in the curated phrase library, helping users find appropriate clips more efficiently and improving perceived controllability.

\customdance~ improves overall authoring efficiency while enhancing dance quality (Table~\ref{tab:study1_efficiency}).
Compared with \textit{w/o Retriever}, \customdance~reduces average authoring time from 28.28 to 11.66 minutes and reduces clip replacements from 20.12 to 5.32.
These reductions are consistent with the role of multimodal ranking in reducing search effort during phrase authoring.
Moreover, conditions with the Diagnoser enabled show zero detector-flagged artifacts related to local kinematic discontinuities, suggesting that Diagnoser-guided repair reduces kinematic anomalies.

\begin{table}[t]
  \centering
  \caption{Results for authoring efficiency and artifacts.}
  \label{tab:study1_efficiency}
  \begin{tabular}{lccc}
    \toprule
    Condition & Time (min)$\downarrow$ & Replacements$\downarrow$ & Artifacts$\downarrow$ \\
    \midrule
    \customdance & \textbf{11.66$\pm$2.58} & \textbf{5.32$\pm$2.56} & 0.00 \\
    Baseline & 28.41$\pm$1.71 & 25.16$\pm$7.41 & 5.56$\pm$1.76 \\
    w/o Anchors+Cues & 13.21$\pm$2.33 & 8.84$\pm$2.97 & 0.00 \\
    w/o Retriever & 28.28$\pm$1.94 & 20.12$\pm$6.77 & 0.00 \\
    w/o Diagnoser & 13.38$\pm$3.18 & 6.08$\pm$2.84 & 5.52$\pm$2.12 \\
    \bottomrule
  \end{tabular}
\end{table}

\begin{figure*}[t]
  \centering
  \includegraphics[width=0.9\linewidth]{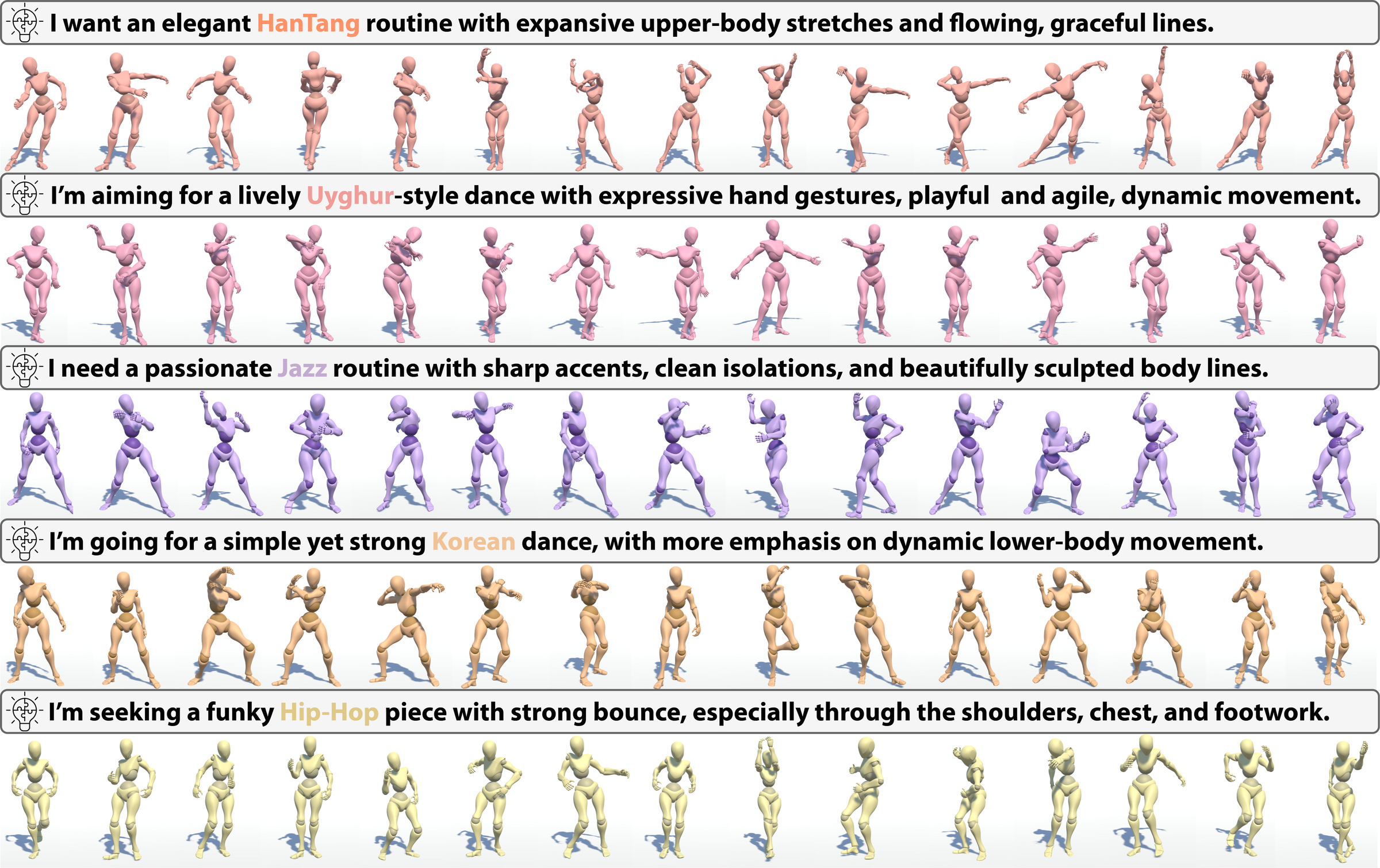}
  \caption[Qualitative results across dance styles]{Qualitative results of CustomDance across five FineDance coarse styles (Classic, Folk, Standard, Mix, Street), conditioned on user preferences.}
  \label{fig:genre}
\end{figure*}

These objective measurements of efficiency and quality are also reflected in participants' subjective ratings. 
We calculated the Pearson correlation coefficient to evaluate the relationship between participants' ratings and our quantitative efficiency and quality metrics.
We found that artifact count inversely correlates with pose satisfaction (Pearson $r=-0.317$, $p<.001$).
Furthermore, increased authoring time inversely correlates with pose satisfaction ($ r =- 0.756$, $p < .001$).
Together, these results indicate that lowering kinematic artifacts and editing burden is directly reflected in participants' subjective evaluations.
Fig.~\ref{fig:genre} shows choreographies authored with \customdance~across five coarse dance-style classes, demonstrating coherent phrase structures and smooth transitions.

\subsection{Study 2: Dance Quality Comparison}
\label{sec:study_quality}

\subsubsection{Dance Quality Comparison Design}\mbox{}\par

\ssection{Participants.}
We conducted a survey involving 25 participants to evaluate the quality of dances created with \customdance~in comparison to recent 3D music-to-dance generation methods.
The participants represented a range of experience in dance performance, with experience levels from 0 to 7 years (M = 3.02, SD = 2.14).
The study was approved by our university IRB, and all participants provided informed consent.

\ssection{Comparison Setup.}
We compare \customdance~with two recent 3D music-to-dance generation methods that improve choreography generation from complementary perspectives: Lodge~\citep{li2024lodge} performs coarse-to-fine diffusion-based long dance generation for global choreography structure and local motion refinement, while MEGADance~\citep{yang2025megadance} uses genre-aware expert modeling with MoE and hybrid sequence modeling to capture genre-specific motion patterns.
\customdance~is designed for interactive choreography authoring: users select, compose, and refine motion phrases according to a music clip and choreography description.
We evaluate Lodge and MEGADance under their native generation regimes and \customdance~under its intended interactive authoring regime.
We compare four conditions: \emph{\customdance}, \emph{Baseline}, \emph{Lodge}, and \emph{MEGADance}.
The \emph{Baseline} is the basic editor from Study~1: it supports timeline-based clip selection but removes motif planning, multimodal retrieval, Diagnoser, and Completer/Remaker.
For a fairer comparison of final choreography quality across these regimes, we then apply the same repair workflow to Baseline, Lodge, and MEGADance before the ranking study, using it only to correct localized artifacts while preserving each method's overall choreography.
The supplementary material details this repair protocol and the artifact-count analysis.

For \customdance~and \emph{Baseline}, we select five dances authored by participants in Study~1 in order to minimize potential biases from raters toward dances created by skilled dancers.
For Lodge and MEGADance, we generate a pool of five dances each using different randomized seeds.
This results in 32-second-long dances synchronized to the same audio for each compared condition.

\ssection{Procedure.}
We employ a blinded multi-alternative ranking procedure where participants first read a choreography description and listen to a corresponding 32-second music clip.
Next, they watch four dance videos (\customdance, \emph{Baseline}, \emph{Lodge}, and \emph{MEGADance}) randomly selected from the pool of dances created using each method with the same music clip and choreography description.
All videos are rendered with identical avatar and camera settings, synchronized to the same audio, with the identities of the methods concealed and the video layouts randomized.

After viewing the videos, raters rank all four dances from best (rank 1) to worst (rank 4) based on three criteria: \emph{Music Alignment} (``Which dance better matches the music?''), \emph{Description Fulfillment} (``Which dance better matches the given description?''), and \emph{Overall Performance} (``Which dance is better overall?'').
Raters are instructed to evaluate the dances solely based on the displayed choreography description and audio, without making assumptions about which methods are capable of accessing the choreography description.
This process is repeated 25 times until the rater has viewed all five dances from each method.

\subsubsection{Dance Quality Results}\mbox{}\par

\begin{figure*}[t]
  \centering
  \includegraphics[width=\linewidth]{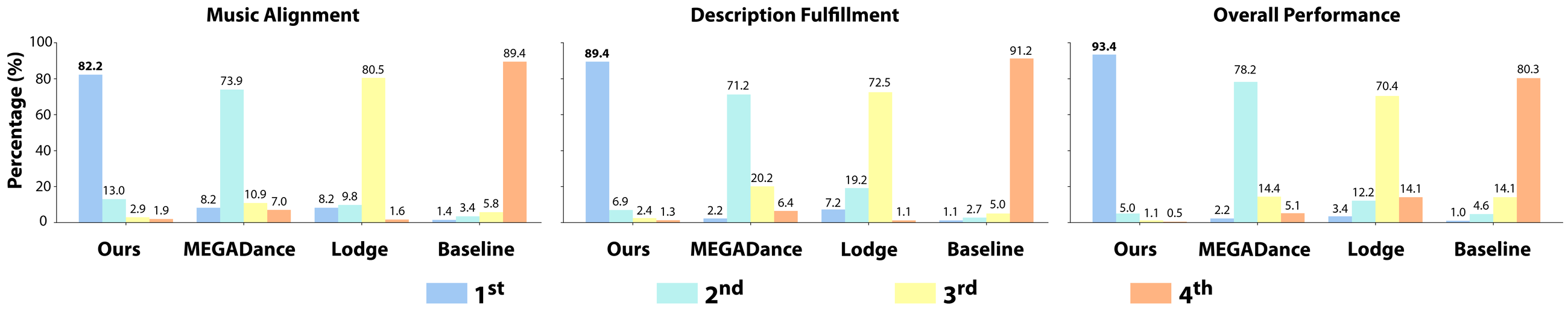}
  \caption[Rank distributions]{Rank distributions of CustomDance, MEGADance, Lodge, and Baseline across Music Alignment, Description Fulfillment, and Overall Performance.}
  \label{fig:winrate}
\end{figure*}

\begin{figure*}[t]
  \centering
  \includegraphics[width=0.9\linewidth]{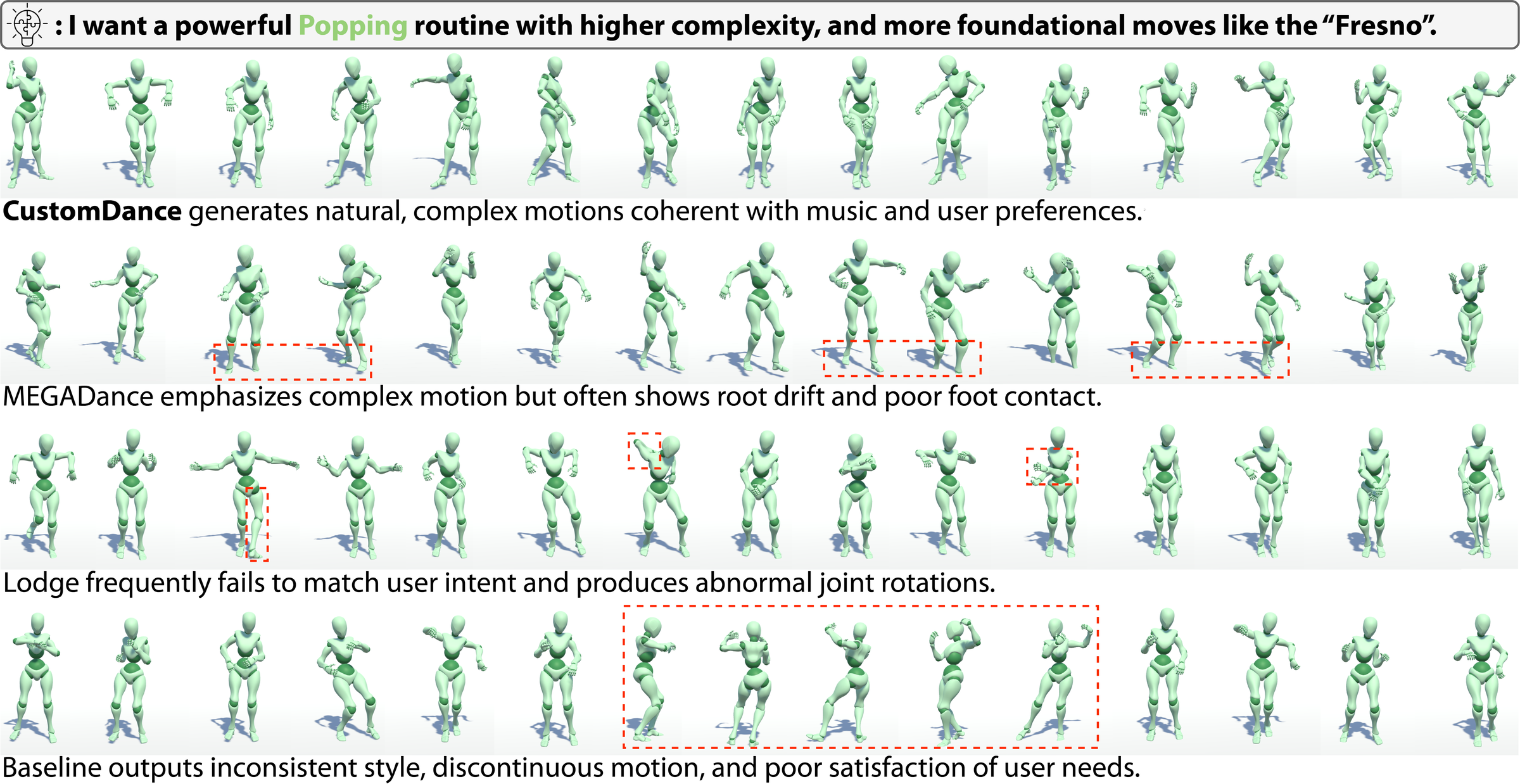}
  \caption[Qualitative comparison]{Qualitative comparison on a street-style excerpt conditioned on user preferences.}
  \label{fig:cmp}
\end{figure*}

\ssection{Subjective Results.}
\customdance~is most frequently preferred across all three criteria, achieving rank-1 rates of 82.2\% (\emph{Music alignment}), 89.4\% (\emph{Description fulfillment}), and 93.4\% (\emph{Overall Performance}) (Fig.~\ref{fig:winrate}). \customdance~demonstrates more coherent phrasing and smoother transitions than the \emph{Baseline} (Fig.~\ref{fig:cmp}).
It also features richer motion and more natural poses compared to \emph{Lodge} and \emph{MEGADance}.
Fig.~\ref{fig:cmp} shows a qualitative comparison between \customdance~and the other methods before artifact processing, illustrating characteristic local artifacts such as abrupt root motion or abnormal joint rotations.

\ssection{Objective Results.}
We report Fréchet Inception Distance and Diversity in kinematic and geometric feature spaces~\citep{li2021ai, siyao2022bailando}, and Beat Alignment Score (BAS)~\citep{li2024lodge} for music--motion synchronization.
Table~\ref{tab:objective} summarizes objective metrics commonly used in music-to-dance evaluation.
\customdance~obtains the lowest FID$_k$ (22.55) and the highest Div$_k$ (7.02), Div$_g$ (6.44), and BAS (0.233) among the compared methods, indicating strong motion fidelity, diversity, and music--motion synchronization.

\begin{table}[t]
  \centering
  \caption{Objective evaluation of dance quality.}
  \label{tab:objective}
  \setlength{\tabcolsep}{3.5pt}
  \begin{tabular}{lccccc}
    \toprule
    Method
    & FID$_k\downarrow$ & FID$_g\downarrow$
    & Div$_k\uparrow$ & Div$_g\uparrow$
    & BAS$\uparrow$ \\
    \midrule
    Baseline
    & 67.23 & 34.13
    & 3.23 & 3.63
    & 0.201 \\
    Lodge [CVPR'24]
    & 35.67 & 24.23
    & 4.32 & 4.57
    & 0.211 \\
    MEGADance [NeurIPS'25]
    & 32.59 & \textbf{16.76}
    & 5.96 & 5.65
    & 0.226 \\
    \customdance
    & \textbf{22.55} & {21.55}
    & \textbf{7.02} & \textbf{6.44}
    & \textbf{0.233} \\
    \bottomrule
  \end{tabular}
\end{table}

\subsection{Study 3: Long-Sequence Choreography Comparison}
\label{sec:study_long_sequence}

\begin{table*}[t]
  \centering
  \caption{Results of the 90-second long-sequence evaluation.}
  \label{tab:study3_long_sequence}
  \small
  \setlength{\tabcolsep}{6pt}
  \begin{tabular*}{\textwidth}{@{\extracolsep{\fill}}lccccc@{}}
    \toprule
    Condition & Time (min)$\downarrow$ & FSR (\%)$\downarrow$ &
    Jitter (km/s$^3$)$\downarrow$ &
    Pose Satisfaction$\uparrow$ & Description Fulfillment$\uparrow$ \\
    \midrule
    Lodge [CVPR'24] & 0.4$\pm$0.2 & 7.02$\pm$0.14 & 0.45$\pm$0.21 & 2.9$\pm$1.1 & 2.2$\pm$0.7 \\
    MEGADance [NeurIPS'25] & 0.3$\pm$0.1 & 15.78$\pm$0.08 & 0.36$\pm$0.17 & 3.8$\pm$1.4 & 4.4$\pm$0.9 \\
    Non-professional scratch & 117.2$\pm$19.3 & 5.87$\pm$0.11 & 0.31$\pm$0.11 & 5.4$\pm$0.8 & 5.1$\pm$1.3 \\
    Professional scratch & 63.6$\pm$8.9 & 5.17$\pm$0.05 & 0.22$\pm$0.08 & 6.7$\pm$0.2 & 6.8$\pm$0.1 \\
    \customdance & 31.2$\pm$6.7 & 5.24$\pm$0.21 & 0.24$\pm$0.14 & 6.1$\pm$0.7 & 6.3$\pm$0.5 \\
    \bottomrule
  \end{tabular*}
\end{table*}

\subsubsection{Long-Sequence Comparison Design}\mbox{}\par

\ssection{Study Design.}
To provide a controlled long-sequence comparison, we evaluated five 90-second music clips, each paired with a fixed choreography description, across five conditions: \customdance, non-professional scratch choreography, professional scratch choreography, Lodge, and MEGADance.
Fifteen participants with prior choreography experience each authored one choreography with \customdance~and one from scratch in a real-world setting on different music clips; music assignment and task order were counterbalanced.
Five genre-matched professional choreographers each created one scratch choreography in a real-world setting, while Lodge and MEGADance each generated one output per clip under their native generation regimes. We captured all real-world scratch choreographies using the Sony mocopi motion-capture system and extracted the resulting motion data in Blender.

\ssection{Evaluation.}
Five independent choreography experts, blinded to condition identity, rated all 45 videos using the same 7-point \emph{Pose Satisfaction} and \emph{Description Fulfillment} items as in Study~1.
All videos used the same avatar, camera, audio, and motion-retargeting and rendering pipeline.
We report authoring or generation time, FSR following Lodge~\citep{li2024lodge}, and Jitter following DNO~\citep{karunratanakul2024optimizing}. Time denotes native generation runtime for Lodge and MEGADance and active authoring time for the other conditions.
The study received IRB approval, and all participants provided informed consent.
Full participant demographics, procedure, and aggregation details are provided in the supplemental.

\subsubsection{Long-Sequence Comparison Results}\mbox{}\par
As shown in Table~\ref{tab:study3_long_sequence}, \customdance{} required less active authoring time than both scratch conditions, received higher expert ratings than Lodge and MEGADance, and achieved FSR and Jitter values close to the professional scratch references.

\subsection{Component-Level Evaluations}
We summarize component-level evidence for the main design choices; full protocols and results are provided in the supplementary material.

\ssection{Slot Duration Selection.}
A duration study supports using $\Delta_{\text{slot}}=4$\,s for phrase-level retrieval: 2-second units often lacked complete motion semantics, whereas 8-second units reduced recombination flexibility by bundling too much movement into each candidate. This setting is also consistent with eight-count choreography units~\citep{vaganova2012basic} and dance music around 120 BPM~\citep{duman2022music}. 

\ssection{Candidate Preview Budget.}
A preview-usability study evaluated recommendation-list sizes of $K=5$, $10$, and $15$. The results support $K=10$ as the best balance between choice diversity and reviewability: $K=5$ restricted candidate choice, whereas $K=15$ substantially increased review time.

\ssection{Anchor--Cue Validation.} 
An expert rating study validates the anchor--cue stage. Gemini-generated anchors receive higher \emph{Slot Appropriateness} ratings than random and beat-based anchors (5.8$\pm$0.3/7 vs. 2.0$\pm$0.5/7 and 4.2$\pm$0.4/7), and the paired cues receive high \emph{Cue Usefulness} ratings (6.1$\pm$0.2/7).

\ssection{Retriever Evaluation.} 
A retrieval benchmark on held-out phrases shows that the full Retriever achieves Recall@5 of 50.0, Recall@10 of 73.0, and Mean Rank of 16.58, outperforming modality-ablated variants. A separate phrase-selection study using precomputed candidate sets shows that participants selected a phrase faster from Retriever candidates than from Lodge and MEGADance candidates (57.8$\pm$8.3\,s vs. 107.8$\pm$15.0\,s and 85.2$\pm$8.5\,s), and reported higher pose satisfaction (6.4$\pm$0.5 vs. 3.2$\pm$0.8 and 4.4$\pm$0.9) and description fulfillment (6.6$\pm$0.5 vs. 2.8$\pm$1.1 and 4.2$\pm$0.8).

\section{Conclusion}
We presented \customdance, a coarse-to-fine human-in-the-loop system that supports customized 3D dance authoring through choreographic motif planning, multimodal phrase retrieval, and diffusion-based completion and refinement.
Our evaluations demonstrate the effectiveness of this workflow for both the authoring process and the quality of the resulting choreographies.

\ssection{Limitations and Future Work.}
While \customdance~achieves strong results in both authoring experience and dance quality, it still has certain limitations.
Fixed-length retrieval phrases and limited library coverage may provide less relevant support for underrepresented styles, while fidelity-oriented metrics may partially favor retrieval-supported outputs.
The workflow also depends on a closed planning API, which may be less reliable for noisy audio or complex rhythms, while automatic refinement does not fully address semantic or stylistic mismatches and contact-related artifacts.
These cases may still require user judgment or phrase reselection.
Future work could broaden the system's representational coverage through variable-length phrases and more style-balanced motion libraries.
It could also improve robustness through reproducible planning models, contact- and collision-aware refinement, and motion-capture or video-based demonstrations for more expressive control.

\bibliographystyle{ACM-Reference-Format}
\bibliography{sample-base-scholar-partial-20260815}

\end{document}